\newif\ifNotes\Notesfalse
 \newif\ifAnon\Anontrue
\newif\ifCamera\Cameratrue

\ifCamera
    \Notesfalse
    \Anonfalse
\fi

\documentclass[letterpaper,twocolumn,10pt]{article}
\usepackage[hyphens]{url}
\usepackage{style/usenix-2020-09}

\usepackage{tikz}
\newcommand*\circled[1]{\tikz[baseline=(char.base)]{
            \node[shape=circle,draw,inner sep=1pt] (char) {#1};}}

\usepackage{caption,float}
\usepackage{pgf} 
\usepackage{svg} 
\usepackage{epsfig}
\usepackage{cite} 
\usepackage{url} 
\usepackage{pgfplots} 
\usepackage{subcaption}
\usepackage{xspace}
\usepackage{booktabs}
\usepackage{enumitem}
\usepackage{tabularx}
\usepackage{comment} 
\usepackage{amsbsy} 
\usepackage{verbatim} 
\usepackage{pifont}
\usepackage[numbers,sort]{natbib}
\usepackage{amsmath}
\usepackage{tablefootnote}
\usepackage{listings}
\usepackage[capitalise,nameinlink,noabbrev]{cleveref}
\hypersetup{
    colorlinks,
    linkcolor={red!50!black},
    citecolor={darkgreen!70!black},
urlcolor={blue!80!black}}

\usepgfplotslibrary{groupplots}

\ifNotes
    \newcommand{\colorcomment}[2]{\leavevmode\unskip\space{\color{#1}#2}\xspace}
    
\else
    \newcommand{\colorcomment}[2]{\leavevmode\unskip\relax}
    
\fi

\definecolor{darkgreen}{rgb}{0,0.65,0}
\definecolor{darkviolet}{HTML}{9400D3}
\definecolor{wildwatermelon}{HTML}{FF43A4}
\newcommand{\taggedcolorcomment}[3]{\colorcomment{#1}{[\textbf{#2}: #3]}}

\newcommand{\yval}[1]{\taggedcolorcomment{darkviolet}{yval}{#1}}
\newcommand{\peter}[1]{\taggedcolorcomment{blue}{peter}{#1}}
\newcommand{\lukasz}[1]{\taggedcolorcomment{orange}{lukasz}{#1}}

\newcommand{\parhead}[1]{\vspace{1pt plus 1pt minus 1pt}\par\noindent\textbf{#1}\hspace{.4em plus .2em minus .2em}}

\def\checkmark{\tikz\fill[scale=0.4](0,.35) -- (.25,0) -- (1,.7) -- (.25,.15) -- cycle;} 

\newcommand{\instr}[1]{{\fontfamily{qcr}\selectfont #1}\xspace}
\creflabelformat{equation}{#2\textup{#1}#3}
\newcommand{\csum}{\ensuremath{c_{\mathit{sum}}}\xspace}
\newcommand{\cout}{\ensuremath{c_{\mathit{out}}}\xspace}

\usepackage{filecontents}

\begin{document}

\date{}

\title{\Large \bf BarraCUDA: Edge GPUs do Leak DNN Weights}

\author{
	\IEEEauthorblockN{P\'eter Horv\'ath\IEEEauthorrefmark{1},
		{\L}ukasz Chmielewski\IEEEauthorrefmark{2},
		Leo Weissbart\IEEEauthorrefmark{1},
		Lejla Batina\IEEEauthorrefmark{1},
		Yuval Yarom\IEEEauthorrefmark{3}}
	\IEEEauthorblockA{\IEEEauthorrefmark{1}Radboud University}
	\IEEEauthorblockA{\IEEEauthorrefmark{2}Masaryk University}
	\IEEEauthorblockA{\IEEEauthorrefmark{3}Ruhr University Bochum}
}

\author{
	\ifAnon\else
	{\rm Peter Horvath}\\Radboud University \and
	{\rm Lukasz Chmielewski}\\Masaryk University, Radboud University \and
	{\rm Léo Weissbart}\\Radboud University\and
	{\rm Lejla Batina}\\Radboud University \and
	{\rm Yuval Yarom}\\Ruhr University Bochum
	\fi
} 

\maketitle

\begin{abstract}
	Over the last decade, applications of neural networks have spread to every  aspect of our lives.
	A large number of companies base their businesses on building products that use neural networks for tasks such as face recognition, machine translation, and self-driving cars.
	Much of the intellectual property underpinning these products is encoded in the exact parameters of the neural networks.
	Consequently, protecting these is of utmost priority to businesses.
	At the same time, many of these products need to operate under a strong threat model, in which the adversary has unfettered physical control of the product.
	In this work, we present BarraCUDA, a novel attack on general-purpose Graphics Processing Units (GPUs) that can extract parameters of neural networks running on the popular Nvidia Jetson devices.
	BarraCUDA relies on the observation that the convolution operation, used during inference, must be computed as a sequence of partial sums, each leaking one or a few parameters.
  Using correlation electromagnetic analysis with these partial sums, BarraCUDA can recover parameters of real-world convolutional neural networks.
\end{abstract}

\section{Introduction}

The field of machine learning has seen an explosive increase in interest and use over the last decade.
In particular, deep learning has proven to be a versatile technique that provides state-of-the-art performance for many real-world applications.
The use of Deep Neural Networks (DNNs) has proved useful for a broad range of domains, including playing chess~\cite{silver2017mastering},
object detection~\cite{liu2020deep},
image classification~\cite{krizhevsky2012imagenet, simonyan2014very, lin2013network, chollet2017xception, he2016deep},
audio processing~\cite{purwins2019deep},
forecasting~\cite{laptev2017time, rangapuram2018deep, sagheer2019time, salinas2020deepar}
and natural language processing~\cite{otter2020survey}.
Thus, deep learning applications have become indispensable and are changing many aspects of our everyday lives.

Deep learning typically
employs artificial neural networks consisting of multiple layers of (simulated) neurons.
When designing a deep learning solution for a problem, the designer first chooses the network architecture, which specifies the layers of neurons, including their sizes, types, and activation functions, as well as how the neurons are connected, i.e., which neurons' outputs are connected to which inputs.
The designer then trains the network, selecting the weights used for each weighted sum and bias values that are added to the sums prior to the computation of a non-linear activation function, such as rectified linear unit (ReLU)~\cite{Fukushima69}.

Training a network for any non-trivial example is a resource-intensive process.
There is a need to curate a specialized dataset of correctly labeled samples that can be used for the training. 
Moreover, the training process often requires days and even weeks of computation on specialized high-performance hardware, such as large quantities of graphical processing units (GPUs), and the whole process requires specialized expertise that is now in high demand.
Thus, to protect owners' IP and to defend against potential attacks, trained models are often considered trade secrets, which should be protected from undue disclosure.

At the same time, there is a substantial market incentive for pushing machine learning to edge devices such as intelligent cameras, autonomous vehicles, and drones.
Consequently, trained models are being deployed under a threat model that allows adversarial physical access to devices and exposes them to side-channel attacks.
Indeed, side-channel attacks against neural network implementations on CPUs have been demonstrated using both electromagnetic side-channel analysis~\cite{batina2019csi} and microarchitectural attacks~\cite{YanFT20}, among others.
Similarly, commercial deep-learning accelerators on FPGAs have also been shown to be vulnerable to parameter extraction via power-analysis~\cite{Gongye2023:SCA-DPU}.
However,
GPUs are the dominant hardware in the world of deep learning due to their performance and the software ecosystem that they provide to implement neural networks.
The CUDA parallel computing platform allows developers to quickly and efficiently deploy DNNs on modern GPUs, whose applications have spread to many areas, from data centers to edge computing.
However, side-channel attacks on GPUs are challenging due to their complexity and inherent parallelism.
So far, attacks on GPU implementations have only succeeded in recovering the network architecture but not the parameters~\cite{chmielewski2021reverse, DBLP:conf/uss/MaiaXLGZ22,horvath2024cnn}.

Therefore, this work focuses on the following research question:

\smallskip
\noindent\textit{{Are proprietary implementations of neural networks on GPU
			vulnerable to parameter extraction using side-channel analysis?}}
\smallskip

\subsection*{Our Contribution}
This work answers the question affirmatively.
We perform side-channel attacks on multiple GPU architectures, such as the Nvidia Jetson Nano~\cite{nano} and Nvidia Jetson Orin Nano~\cite{orin_soc}, recovering the weights and biases of real-world networks.

Specifically, our attacks collect multiple traces with random inputs and
then use correlation electromagnetic analysis (CEMA)~\cite{brier2004correlation} adjusted to recover the model's  parameters from these traces.
We further demonstrate the success of our attack on different neural network layers, such as convolutional and dense layers with varying batch sizes.

Our attack relies on the observation that convolution computation, the core operation in many neural network implementations, cannot be computed at once.
Instead, convolution is computed as a sequence of partial sums, each depending on the previous sum and a small number of weights and inputs.
Consequently, assuming the previous sum and the inputs are known, the adversary can guess the weights, use these to predict leakage, and correlate the prediction with the measured side-channel trace.
However, to carry out the attack, we
need to overcome several challenges.

\parhead{Partial Sums Identification:}
Convolutional and dense layers in neural networks can be implemented in multiple ways; one of the most commonly used ways is matrix multiplications~\cite{chellapilla:inria-00112631}.
In a differential SCA attack, knowing how the target algorithm is implemented is crucial.
Therefore, we first reverse-engineer CUDA binaries produced by TensorRT to determine both the parameter representation, the relevant partial sums, and the weights they depend on.

\parhead{Attack Localization:}
A second challenge is that for an effective attack, the attacker needs to localize the attack, both spatially and temporally.
To overcome this challenge, we adapt techniques from the domain of side-channel attacks on cryptographic implementations.
We use a variant of Test Vector Leakage Assessment (TVLA)~\cite{schneider2015leakage} both for finding the best physical location for placing the probe and for finding the time during the operation of the neural network in which a specific neuron is evaluated.
Our leakage detection analysis proved effective for different GPU architectures. \peter{Here I mean the chip and capacitors for localization and HW and HD model for leakage modeling}

\parhead{GPU architecture:}
The GPU inherently introduces a high amount of noise due to its Single-Instruction-Multiple-Thread (SIMT)~\cite{simt} architecture, where many parallel threads are executing simultaneously.
Furthermore, due to the GPU hardware implementation, the scheduling of threads is not guaranteed to be fixed in time, which makes SCA attacks (e.g., CEMA) much harder than on other platforms, such as FPGAs and microcontrollers, where the execution is deterministic and sequential.
Moreover, unlike attacks on cryptographic implementations, where the attacker typically aims to recover a relatively short sub-key, we need to extract a large number of parameters for DNN model extraction.
A large number of traces and efficient signal processing are required to overcome these challenges.
To that end, we develop a CUDA-based implementation of CEMA to execute the attack an order of magnitude faster for large datasets with millions of traces.

\parhead{Parameter Extraction:}
We apply our techniques to convolutional and dense layers in neural networks while also investigating the impact of batch size on our attack.
Our attacks against convolutional layers target a variation of the baseline EfficientNet~\cite{tan2019efficientnet} model.
We deploy the model on the Jetson Nano in FP16 precision and the Jetson Orin Nano in INT8 precision for the parameters.
We successfully and efficiently extract parameters from both GPUs with different data types using our CUDA-based analysis tool.
Overall, the attack against the Jetson Nano requires 10 days for collecting traces and one to two days for trace alignment with an input batch size of 1.
On the other hand, the Jetson Orin Nano requires only 1 day for trace collection and one day for trace alignment with an input batch size of 1.
An experiment with a larger batch size was also conducted on the Orin Nano, and it took 5 days to collect traces and align them with an input batch size of 16.  
Once aligned, parameters can be extracted at a rate of one weight in six minutes for FP16 and 5 minutes for INT8 weights.
The whole process is highly parallelizable.
Thus, attacks on moderate-size models are well within the capabilities of well-resourced adversaries.

\parhead{Disclosure.}
We notified Nvidia of the vulnerabilities found, and they acknowledged our findings. Nvidia recommends that users follow guidelines to prevent physical access and information leakage.

\parhead{Organization.}
The rest of this paper is organized as follows:
After providing the necessary background on side-channel attacks and deep learning (\cref{sec:background}), we describe the overview of our attack and our experiment setup in \cref{sec:leakage_detection}.
In \cref{sec:profiling} we describe our procedure for profiling the implementation, identifying relevant partial sums, and localizing their leakage.
Finally, we present our parameter extraction attack in \cref{sec:pex}.
In addition, \cref{sec:disc} covers the limitations, possible extensions, countermeasures, and related work.

\section{Background}\label{sec:background}

Here, we introduce the concepts of side channels and related techniques used for the attacks, and we provide some background on Nvidia's CUDA programming model and GPU architecture.

\subsection{Deep Neural Networks}
Deep neural networks (DNN) are universal function approximators~\cite{HORNIK1989359} that solve tasks by learning from data.
First, a model learns from training data; then it can be deployed to make predictions on new, unseen data.
Two main components influence the trained model's performance on unseen data: the architecture and the parameters.
The \emph{architecture} refers to the structure of the model, the types, and order of transformations that the model applies on its inputs to arrive at some output.
These transformations are also commonly called \emph{layers}, and their output can be tweaked by changing their internal \emph{parameters}.
During the training process, these parameters are tweaked so that the final output yields correct results.

\noindent \textbf{Convolutional layer.} A convolutional layer consists of small \emph{kernels} that extract different features from the layer's input e.g., an image.
The dimensions of these kernels are usually much smaller than the input's dimensions to be able to extract fine-grained details.
Each kernel extracts different features with its own parameters: the weights and bias.
Each of them calculates the convolution between their parameters and a small part of the input by sliding every kernel on the input with a certain step size.
In this work, we demonstrate the extraction of parameters, the weights, and the bias from convolutional layers.

\noindent \textbf{Dense layers.} A dense layer in a neural network consists of nodes where each node outputs the weighted sum of all the inputs. 
In this paper, we demonstrate the extraction of weights from dense layers.

\subsection{Side-Channel Analysis}

Side-Channel Analysis (SCA) exploits unintended physical leakages of electronic devices to extract secret information processed by them~\cite{kocher1996timing, kocher1999differential}.
Such leakage can occur through various channels, including power consumption, electromagnetic emanations (EM), timing, optical, or sound. It might lead to leakage of various types of secret information, e.g., on data and instructions processed.
In academic settings, Side-Channel Analysis (SCA) was first introduced in the 90's, targeting cryptographic implementations running on then popular but constrained cryptographic devices such as smart-cards~\cite{kocher1996timing, kocher1999differential} and SCA poses ever since a constant threat to the security of various embedded systems.
In that context, SCA aimed to recover the secret keys used in the cryptographic implementations.
In this work, we exploit the EM side channel emanating from a GPU platform on which a neural network is running, but instead of targeting secret keys, we show how to recover neural network secret parameters: weights and biases.

\peter{I think the conv and dense layer descriptions can be omitted if needed.}

\subsubsection{Electromagnetic Emanation}
The electromagnetic
emanations from a computing device correlate with the code and data the device processes.
This correlation has been used to break cryptographic implementations~\cite{QS01, kuhn1998soft}, reverse-engineer neural networks~\cite{batina2019csi, chmielewski2021reverse} and eavesdrop on display units~\cite{elibol2012realistic, hongxin2009recognition, liu2020screen}.


In Correlation EM Analysis (CEMA)~\cite{brier2004correlation}, the attacker uses the correlation coefficient as a side-channel distinguisher, i.e., the statistical method used for the key recovery.
Essentially, CEMA allows an attacker to recover parts of a secret that
is used in a targeted operation by using a known plaintext attack: measured samples are correlated against a synthetic leakage value (i.e., leakage model) that is generated from an intermediate value calculated for all possible values of a (part of) the secret. 
In our case, all CNN parameters (i.e., weights and bias) are considered as the whole secret, and a single weight or bias is considered to be a single target of CEMA that needs to be repeated for all of them; the intermediate values are results of computations within neurons.
Observe that for the above approach to work, we need to assume that the inputs used in the computations are different and are known to the attacker.

Two commonly used leakage models in SCA are the Hamming weight (HW) model, which predicts that the leakage is linear with the number of set bits in the data (i.e.\ its Hamming weight), and the Hamming distance (HD) model, which predicts that the leakage is linear with the number of bits that flip between consecutive data values. 
HW leakage usually occurs in practice when a value is transferred via the system bus, and HD leakage occurs when an intermediate value stored in a register is overwritten with another value.
We consider both of these models in this paper.

\subsubsection{Leakage Assessment}\label{sssec:leak}

For leakage assessment, which is a critical part of a security evaluation of a chip, we rely on the default techniques used in side-channel analysis, i.e., intermediate-value correlation and Test Vector Leakage Assessment (TVLA)~\cite{schneider2015leakage}.
The idea behind intermediate-value correlation is to consider the correlation traces generated for all possible values of the secret. There should be a correlation peak in the trace when the correct guess (for the secret) is processed because the guess is then in agreement with the leakage predicted. 
This is a consequence of the fact that processing different data causes different physical information, such as timing, power, or EM, often referred to as leakage.
While this approach is often used in actual side-channel attacks, the disadvantage is that it requires a large number of traces, similar to the CEMA attack. 
TVLA or other leakage detection methods, like ${\chi}^2$-test~\cite{moradi2018leakage}, are often more efficient (faster) as we control the parameters in the leakage detection phase.
The main idea of TVLA is to check whether two distributions that process different intermediate values are equivalent or not using Welch’s t-test. 
If the two groups are deemed equivalent, then the TVLA will not observe the leakage. 
Concretely, we verify whether two sets of measurements show significant differences if one set has a
fixed weight and the other has random weights. We refer to this setting as “fixed versus random”.  Since TVLA searches for any leakage not necessarily exploitable
by an attack like CEMA, it also usually requires much fewer traces
than intermediate-value correlation. Therefore, for the sake of
efficiency, we mainly use TVLA in this paper, especially for
determining which part of the traces processes each weight using
TVLA. We use intermediate-value correlation only to
determine the best location of our probe\footnote{We simply check at
which location the absolute t-test peak is the highest as in~\cite{Danial2019SCNIFFERLA}.} and for preliminary
characterization of leakage patterns. 

\subsection{CUDA Programming Model}
To leverage the
parallelism offered by GPUs, Nvidia exposes the CUDA programming model~\cite{cupm} to developers.
In this model, multiple abstraction levels exist and each level has different implications 
forGPU hardware.
The lowest level of abstraction is the
	{\fontfamily{qcr}\selectfont
		thread}, which executes a CUDA function defined by the developer.
The number of threads executing the CUDA function in parallel is specified at the time of invoking the function.\footnote{Functions in CUDA are also called \emph{kernels}. We use the term function to avoid confusion with kernels in CNNs.}
Subsequently, multiple threads can be grouped together into a single
	{\fontfamily{qcr}\selectfont
		block} of threads. Threads in a block have a per-block on-chip shared memory region where they can exchange data with other threads. 
Blocks of threads form a
	{\fontfamily{qcr}\selectfont
		grid} of thread blocks. Each block in a grid executes independently from other blocks, but all blocks in the grid share the same off-chip global memory region.


\subsection{GPU Streaming Multiprocessor}

\begin{table}[t]
	\centering
	\begin{tabular}{lll}
		\toprule
		                 & Jetson Nano & Jetson Orin Nano \\ [0.5ex]
		\midrule
		GPU architecture & Maxwell     & Ampere           \\
		\# SMs           & 1           & 8                \\
		\# CUDA cores    & 128         & 1024             \\
		\# Tensor cores  & -           & 32               \\

		Max. clock       & 920 MHz     & 625 MHz          \\
		FP16             & \checkmark  & \checkmark       \\
		INT8             & -           & \checkmark       \\

		\bottomrule
	\end{tabular}
	\caption{Comparison between the GPUs of Jetson Nano and Jetson Orin Nano.}\label{table:maxwell_vs_ampere}
\end{table}

When a CUDA function is invoked, the parallel threads execute on the GPU's Streaming Multiprocessor (SM). 
A GPU can consist of one or more SMs to improve parallelism further. 
In this paper, we mount our attack on two different GPU architectures, the Maxwell~\cite{tx1} and Ampere~\cite{orin_soc} GPUs embedded in a System-on-Chip (SoC).
For both architectures, when blocks of threads are scheduled onto a particular SM, the threads in the blocks are
divided into groups of 32 threads, also called {\fontfamily{qcr}\selectfont
		warps}. Every warp is assigned to a particular Processing Unit (PU) in the SM, and the warp scheduler in a PU is responsible for scheduling and issuing instructions for warps that are ready for execution at every clock cycle.
Additionally, both architectures have 4 PUs per SM, each with dedicated resources (e.g., register file) to manage warps. This means that 4 warps can be issued instructions in parallel at every clock cycle in an SM.
The Jetson Nano and Jetson Orin Nano's GPUs are similar, but significant differences are summarized in \cref{table:maxwell_vs_ampere}. 
Overall, the Ampere GPU is a larger and more capable GPU with more hardware support for different operations (matrix-multiply) and data types (e.g. INT8) that are heavily used in deep-learning inference.

\subsection{TensorRT Workflow}

TensorRT is a framework dedicated to accelerating neural network inference on GPU, using implementations from different libraries (CuDNN, CuBLAS, TensorRT) that are timed against each other to choose the fastest. We use the TensorRT framework in our experiments and demonstrate the parameter extraction attack on the implementations provided by the framework.

\section{Attack Procedure}\label{sec:leakage_detection}

\begin{figure}[tb]
	\centering
	\includesvg[width=\linewidth]{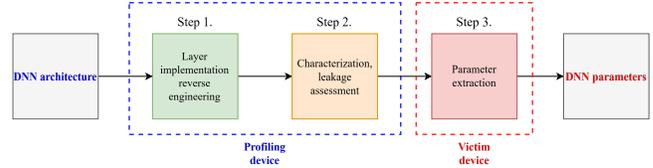}
	\caption{High-level description of the attack procedure.}\label{fig::attack_proc}
\end{figure}

Our attack, depicted in \cref{fig::attack_proc}, assumes that the adversary knows or can recover~\cite{chmielewski2021reverse, DBLP:conf/uss/MaiaXLGZ22, horvath2024cnn}, the DNN model architecture. 
The adversary aims to use side-channel observations to recover the unknown weights and biases used in the victim device.
The attack follows two main phases.
In the first, the adversary uses their prior knowledge to instantiate an equivalent DNN architecture on a profiling device with identical hardware, albeit without the weights.
This profiling device is used to learn how information about weights and biases leaks.
This knowledge is then used to guide the second phase of the attack, which recovers the desired information.

The attack builds on the observation that convolution is the core operation performed during inference.
Specifically, 
A single $p \times q$ kernel with weights $w$, bias $b$ and input feature map $x$ calculates as follows:
\begin{align}\label{eq:csum}
	 & \csum = \textbf{x} * \textbf{w}  = \sum_{i=1}^{p \cdot q} w_i \cdot x_i = w_1 \cdot x_1 + \dots + w_{p\cdot q} \cdot x_{p\cdot q}, \\
	 & \cout = \csum + b,
\end{align}
where $\csum$ is the result of the convolution between the weights and the input, while $\cout$ is the output of the kernel (if the kernel contains a bias).
Finally, if the layer includes an activation function $f$ then it is applied after the convolution:
\begin{align*}
	 & c_{f} = f(\cout).
\end{align*}

Since the convolution in \cref{eq:csum} cannot be computed in a single step, its computation consists of a series of intermediate computations, 
each depending on one on a small number of weights. 
Our attack targets these intermediate computations, looking for correlations between the results of the computations 
and the side-channel observations. 
We refer to these results as \emph{intermediate values}. 

The profiling phase aims to \emph{identify} the intermediate values used in the implementation and to \emph{localize} at which time points they leak in the traces.
The attack then performs correlation analysis on the side-channel information at the identified locations to recover the weights.

The rest of this section outlines our threat model and describes the experimental setup.
The attack itself is described in two sections.
First, \cref{sec:profiling} describes the profiling phase.
Then, \cref{sec:pex} describes and evaluates the recovery of weights and biases from the victim device.

\subsection{Threat Model}\label{ssec:threatmodel}
Our attack targets edge devices that execute machine learning inference for their functionality.
The attacker aims to recover the trade secrets encoded in the parameters (weights and biases) of the machine learning model, for example, in order to steal the IP that encodes them or as a step in designing an adversarial attack on the machine learning model.

We assume that the target device operates correctly and that the code is secure so the attacker cannot exploit programming vulnerabilities to acquire the parameters.
However, we assume that the attacker knows the architecture of the model, including the number of layers, their sizes and types, and how they interconnect.
Attackers that do not have the information can use techniques developed in past works to recover the architecture~\cite{chmielewski2021reverse, DBLP:conf/uss/MaiaXLGZ22, horvath2024cnn}.

As is typical for edge devices, we assume the attacker has unfettered physical access.
In particular, we assume the attacker can open the device and place electromagnetic probes at locations that leak information.
As we discuss in \cref{ssec:setup}, our target devices have multiple leaky locations, allowing the attacker a choice.

Last, we assume that the attacker can monitor the electromagnetic emanations from the target device during the time that the device performs inference.
The attacker needs to be able to observe the emanations over multiple sets of inputs and should also be able to choose these inputs if the target parameters have INT8 data to reduce complexity. 
Otherwise, FP16 parameters do not require control of the inputs.

\peter{Disclaimer: due to chosen input, I changed the last sentence of the threat model. Please review.}
\lukasz{Is it really true? Can the attacker just take more traces and then disregard many? Or attack 32 bits?}


\subsection{Sensitive Intermediate Values}
A successful CEMA attack against a target algorithm requires the attacker to find sensitive intermediate values that depend on secret data. 
In convolutional and dense layers, a sensible choice for these intermediate values is the \emph{partial sums} that depend on the secret weights.
The partial sums allow an adversary to attack one or a few weights at a time and, therefore, to reduce complexity. If an attacker targeted the final results of these layers,
the complexity would increase as these results depend on many secret weights, making the attack infeasible.

However, the actual \emph{implementation} of these layers, and subsequently the computation of partial sums, are dependent on the target layer and hardware characteristics.
Some of the important layer characteristics are the number of input and output channels in a layer and the used representation for the parameters.
In addition, hardware characteristics such as the number of SMs and shared and global memory size can also influence this implementation.
Consequently, different layer and hardware configurations can lead to slightly different implementations. Although these implementations can have identical structures, fine-grained details such as partial sum computations can differ.
Therefore, an attacker first needs to reverse engineer how each layer of the target DNN architecture is implemented on the target device. 
This task can be aided via reverse-engineering of the GPU framework and analysis of leakage of the profiling device.\footnote{One can also imagine an attacker without a profiling device searching through all the possible intermediate values to find the correct one. This approach is feasible but time-consuming, and we do not follow it.} 

\subsection{Experimental Setup}\label{ssec:setup}

To gather side-channel information, we collect EM traces as they are less invasive and can provide more localized information than power measurements. It is also closer to the real world as fewer modifications to the chip are required.
Our two targets, the Jetson Nano and Jetson Orin Nano, are similar in that both feature a SoC mounted onto a PCB in a flip-chip package.
However, the Jetson Orin package is significantly larger than the Jetson Nano's while also surrounded by more capacitors. In order to access the packages, we remove the heatsinks from both devices and use an external fan to provide cooling to the devices.
In our setups, the  GPU cores operate at the highest supported clock frequencies as shown in \cref{table:maxwell_vs_ampere}.
We use the Lecroy 8404M-MS oscilloscope at a sampling rate of 10GS/s with a Langer MFA-R 0.2\nobreakdash-75 near-field probe~\cite{langer_probe} to collect electromagnetic traces.
In our experiments, we find that we need a sampling rate of at least 5\,GS/s to see leakage.
Increasing the frequency up to 20\,GS/s does not improve the results.
In the FP16 case, the number of samples in a single trace for the first convolutional layer is 400\,000 with a batch size of 1.
In the INT8 case, it is 150K per trace.

\begin{figure}[t]
    \centering
    \begin{subfigure}[b]{0.2\textwidth}
        \centering
        \includegraphics[width=\textwidth]{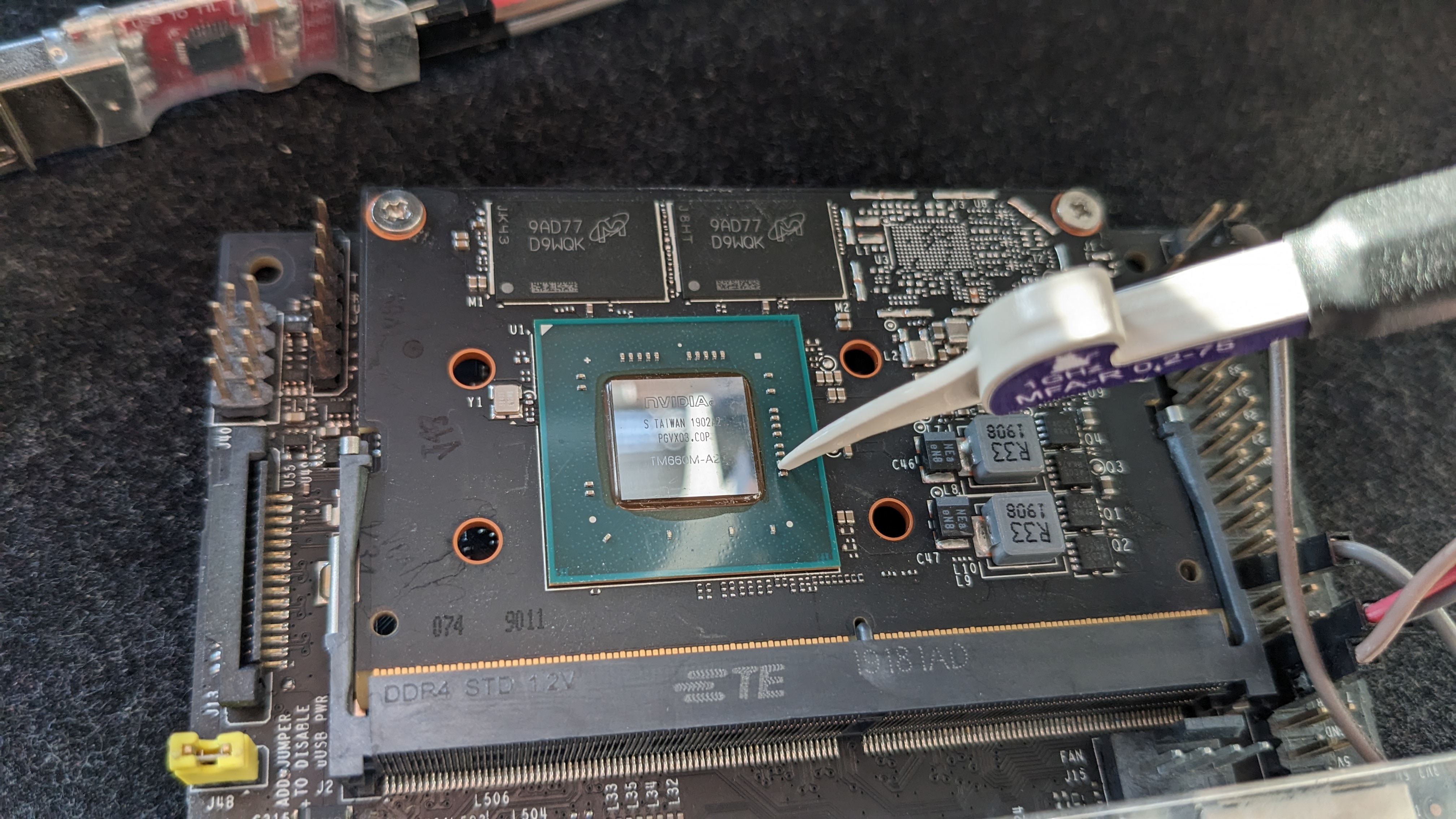}
        \caption{Location of Langer EM probe between two capacitors on the Jetson Nano.}
    \label{fig::setup_jetson_nano}
    \end{subfigure}
    \hfill
    \begin{subfigure}[b]{0.2\textwidth}
        \centering
        \includegraphics[width=\textwidth]{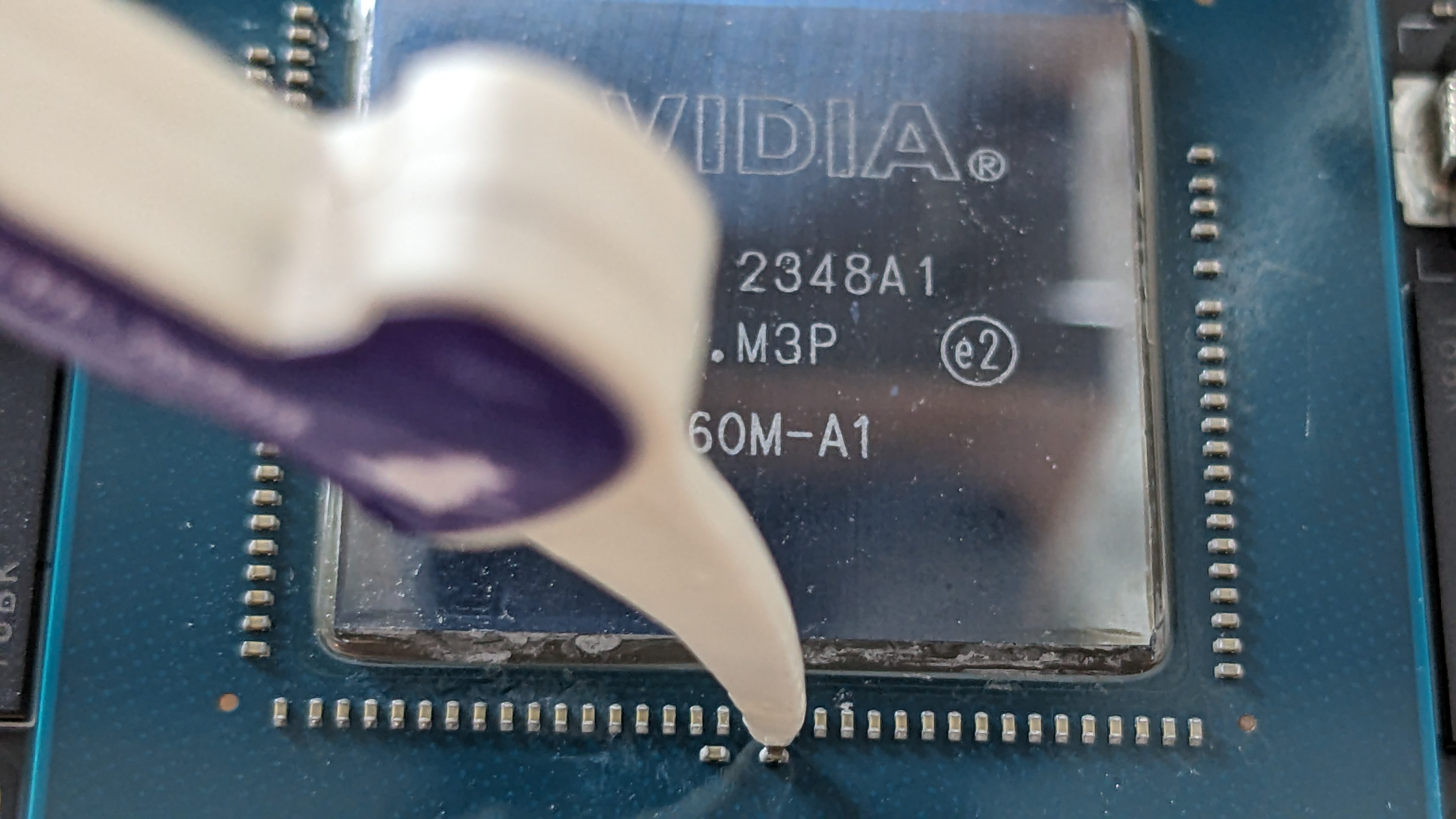}
        \caption{Location of Langer EM probe between 3 capacitors on Jetson Orin Nano.}
    \label{fig::setup_orin_nano}
    \end{subfigure}

	\caption{EM probe locations for the Jetson Nano and Jetson Orin Nano devices for successful parameter extraction attacks.}
	\label{fig::probe_setups}
\end{figure}

\subsubsection{EM Probe Positioning}
To find the best position for the EM probe, we use both TVLA and intermediate-value correlation experiments.
Specifically, we instantiate models whose architectures are identical to the target model for each possible location.
We then capture two sets of traces.
The weights and inputs are always the same in the ``fixed'' set of traces.
In the ``random'' set of inputs, we select a random value for one of the weights in the model, and the inputs are the same. 
We then use TVLA to measure the statistical difference, expressed as the $t$-value,  between the set of traces and intermediate-value correlation to find the correlation between the random weights and the leakage.
Significant correlation and $t$-value peaks indicate a strong signal.

\parhead{Jetson Nano}
We observe several promising locations for placing the EM probe.
\cref{fig::setup_jetson_nano} shows one such location between two capacitors in the power-supply circuit of the Jetson Nano.
Additionally, using XY scan, we find several locations that exhibit similar leakage on the surface of the Jetson Nano's SoC.
Parameter extraction is possible from both the best location on the SoC and between the capacitors.
In the rest of this paper, we use the capacitors' location because probe placement is easier and does not require a detailed scan.

\parhead{Jetson Orin Nano}
On the Jetson Orin's chip surface, we cannot pick any GPU-related signal. 
However, some nearby capacitors still leak
information related to GPU activity.
\cref{fig::setup_orin_nano} shows one vulnerable spot that allows for parameter extraction. 
With further experimenting over the chip surface, we find that probes sensitive to higher frequencies, such as the Langer RF\nobreakdash-B 0.3-3~\cite{langer_probe_rf}, can pick up exploitable signals over the chip as well.
In this paper, the presented results use the location shown in \cref{fig::setup_orin_nano} with the Langer MFA\nobreakdash-R 0.2-75 probe.  

\subsubsection{Trace Acquisition}
In order to extract the parts of the traces related only to the inference operation, we used \instr{nsys} from the CUDA Toolkit to get information about
the execution times of the operations on the GPU.
Therefore, we used the Lecroy oscilloscope's SmartTrigger feature to trigger on the rising edge of the first layer as it proved to be more reliable.

\subsubsection{Trace Preprocessing}\label{ssec:align}

In general, the collected traces contain a lot of jitter and the clock of the cores is not stable, which can be confirmed by looking at the traces in the frequency domain.
This makes the detection of time where the implementations leak and the subsequent CEMA attack harder.
Since aligning the traces accurately with static alignment~\cite{mangard2008power}\footnote{Static alignment employs a standard pattern-based approach: we select a part of a trace as a reference and compute correlation for each offset within a chosen range for each of the traces. We then shift each trace by the respective offset that maximizes the correlation.} at many locations at the same time is not possible,
we use elastic alignment~\cite{van2011improving}\footnote{Elastic alignment is a parametrized machine-learning-based technique to align the traces on all the distinctive patterns at the same time. 
However, this method 
tends to be error-prone and results in a decreased leakage in practice.} 
to improve the leakage detection process.
Although elastic alignment performs alignment at every time point, it comes at a price: it requires finding the optimal input parameters and is computationally expensive.
Moreover, despite tuning the 
parameters, it decreases the amount of leakage in the traces.
Therefore, it is used only to 
detect leaking points, but the CEMA attack is carried out on the raw traces after static alignment is applied on the leaky part of the trace.

\section{Profiling}\label{sec:profiling}

We now turn our attention to the profiling phase of the attack, in which the adversary uses a profiling device to analyze the model and identify trace positions in which partial sums leak.
Profiling consists of two main steps.
In the first step, the adversary analyzes the software that the GPU executes to identify partial sums that depend on weights and characterize the dependencies.
Once these are found, in the second step, the adversary collects side-channel traces from the profiling device and uses statistical tools to find the leaking time points in the traces at which leakage of each weight and bias can be observed.

\subsection{Layer Implementation}\label{ssec:layerimp}
In this section, we discuss the high-level structure of the code that performs the operations of convolutional and dense layers in DNN models.
As the code is unavailable to us, we use \instr{cuobjdump}~\cite{cuobjdump} to produce assembly code, which we can analyze. 
While each implementation is different, all convolutional implementations  follow the same structure:
\begin{itemize}[nosep, leftmargin=*]
\item\circled{1} block of initialization instructions,
\item\circled{2} block of convolution operations, and
\item\circled{3} block of bias addition and ReLU calculation.
\end{itemize}
We now explain these three blocks in more detail.

\parhead{\circled{1} Init Block.}
The first main block consists of instructions that set up the CUDA function.
This block initializes the 64  accumulator registers  that are later used to store the convolution results (R0--R63).
Higher registers (R64 and above) are used to load the weights and inputs.
Since the GPU registers are 32-bit wide, each higher register is loaded with either two FP16 values or four signed 8-bit integers, depending on the data type used for the computations.

\parhead{\circled{2} Convolutional Block.}
The second block performs the convolution of the weights and the inputs, i.e., the partial sums are computed in this block. 
It consists of repeated vectorized loads and arithmetic instructions. 
A set of higher registers is assigned to be loaded with inputs and weights from different input channels.
In addition, this block is executed multiple times depending on the hyperparameters of the convolutional layer, such as the kernel size.
\yval{Needs some more discussion}\peter{Added more.}

\parhead{\circled{3} ReLU Block.}
The third block adds the biases and performs the activations.
For the FP16 implementation, the two partial sums of the accumulator need to be combined before adding the bias and calculating the ReLU.
Conversely, in the INT8 implementation, the partial sum results are already summed into a single register, but there is a need to convert this number to a floating-point prior to adding the bias and applying ReLU.
Unlike the FP16 implementation, which uses half precision throughout the computation, the INT8 implementation converts the integer to a single-precision floating-point number.

\begin{figure}[htb]
	\centering
	\includesvg[width=\columnwidth]{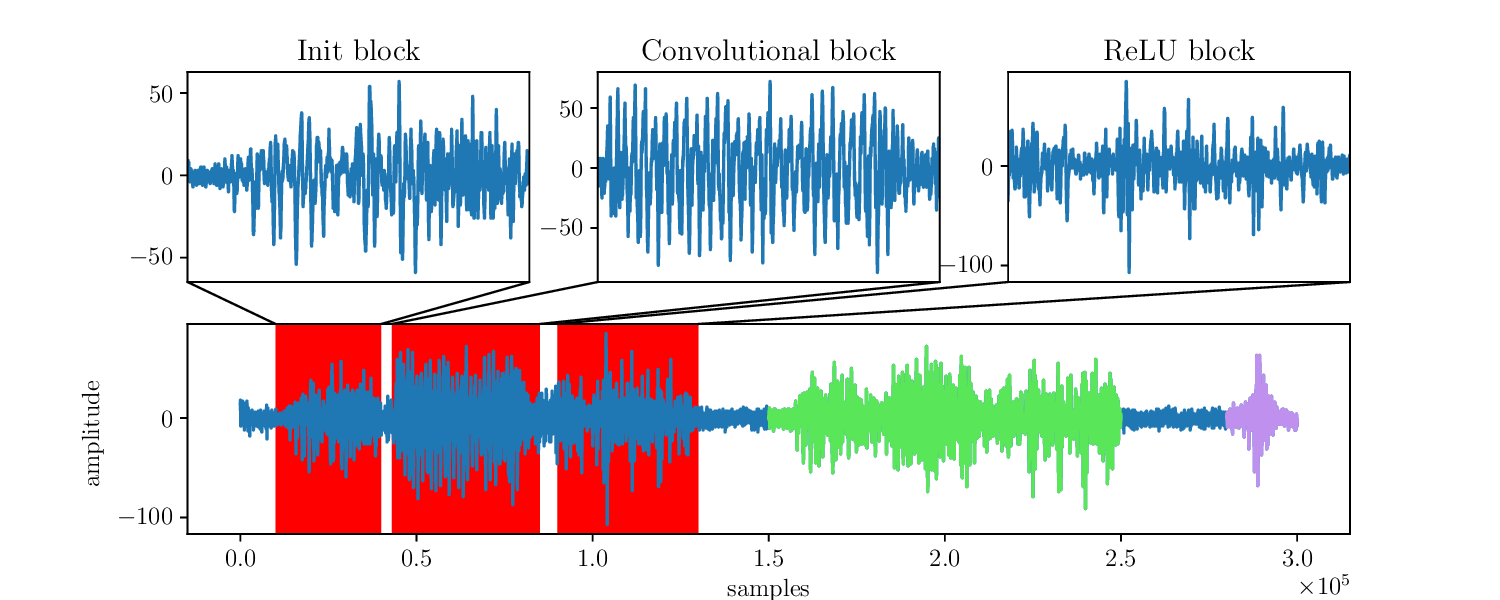}
	\caption{Raw trace of the whole operation on the GPU of Jetson Nano. The two convolutional layers (light pink and yellow) are clearly separated in the traces. Additionally, the CUDA device-to-host memory copy (light purple) is also clearly visible in the end of the trace.\lukasz{maybe mention which color marks which layer?} \peter{fixed}}\label{fig::ex_full}
\end{figure}

\parhead{Matching Instruction Block to Traces.}
\cref{fig::ex_full} shows the electromagnetic emanations of the Jetson Nano GPU during the execution of a CNN with two convolutional layers. 
Since each layer is mapped to a separate CUDA function call by the framework, it can be observed that there is a clear separation between layers.
Additionally, the convolution results are copied back to memory, which is a separate CUDA call and is visible in the trace.

In addition, the three blocks of the implementations can also be identified. 
The first highlighted part shows the init block in the first layer.
The second highlighted block corresponds to the convolutional block where the partial sums are computed. The third highlighted segment corresponds to the calculation of the bias addition and ReLU output, repeated four times for different sets of registers.
Note that these instruction blocks are also separated by synchronization instructions, which are also visible in the trace as the amplitude of the EM signal
drops close to 0 between the blocks.

\subsection{Identifying Partial Sums}\label{ssec:gpure}
As no single GPU instruction can process an arbitrary number of arguments, the code typically processes these convolutions as a sequence of partial sums.
An example of a naive way of computing \cref{eq:csum} is by sequentially computing partial sums $s_j$ using the formula:
\begin{align}\label{eq:partialsums}
	 & s_j = \sum_{i=1}^{j} w_i \cdot x_i = s_{j-1} + w_{j} \cdot x_{j} & (j \leq p \cdot q)
\end{align}

With this formula, if we assume we know $s_{j-1}$ and $x_j$, we can guess $w_j$ and use the guess to compute $s_j$. 
We then apply a leakage model to the computed $s_j$, e.g., the Hamming distance between $s_{j-1}$ and $s_j$, and search for correlations between the model and observed leakage. 
A high correlation indicates that the guess of $w_j$ is correct.

The challenge is that there are multiple ways in which the code can compute \cref{eq:csum}. 
In particular, the implementation can change the order of computing the sum, and it can also use vector operations to combine multiple additions into a single operation.
As our attack relies on knowing the previous partial sum to guess the weight, we must know how the sum is computed.

In this section, we delve into the implementations of convolutions with two reduced-precision implementations used by the TensorRT framework: INT8 and FP16.
Reduced-precision implementations, such as these, are typically used during inference because they reduce both the latency and the memory use compared to single-precision implementations, which are typically used in the training phase.
Our choice of INT8 and FP16 targets the larger and more challenging parameter sizes in reduced-precision techniques~\cite{NIPS2016_d8330f85, micikevicius2022fp8, quinn-ballesteros-2018-pieces, quinn-ballesteros-2018-pieces, narang2017mixed}, and covers both integer and floating-point parameters.
Here we describe the high-level design of the implementations. 
See \cref{sec:convimpdetails} for further technical details.

\parhead{FP16 Convolution.}
The FP16 implementation uses the \instr{HFMA2} instruction, which performs two half-precision fused-multiply-adds in parallel.
It operates on 32-bit registers, each holding two half-precision floating-point numbers.
It first computes the products of two pairs of numbers in matching halves of two registers and then adds the results to the matching halves of a third register, which acts as an accumulator.
This, basically, splits the convolution computation across the two channels, which are summed at the end.
Our leakage model targets each partial sum (16 bits) that is written into this accumulator register.

\parhead{INT8 Convolution.}
The INT8 implementations of convolutional and dense layers use the \instr{IDP.4A} instruction to perform a 4-way dot product and accumulation operation, depicted in \cref{fig::idp4a}.
The instruction first multiplies the elements in matching channels in two registers.
It then sums the results and adds them to a third register, which stores a signed 32-bit accumulator.

In many cases, not all four channels of the dot product contain data.
Specifically, when the number of input channels is three, the channels of a single input point are convolved with the matching weights.
On the other hand, when the input consists of a single channel, two input points are convolved during each operation.
Values for channels that are not used are set to zero and do not affect the result of the instruction.

Unlike the \instr{HFMA2} instruction, the sum depends on all of the weights that are convolved by the \instr{IDP.4A} instruction. 
Thus, the attack needs to target all of the weights that are used by a single instruction.
\lukasz{is the beginning of the paragraph missing?}\peter{I think it was only just not capitalized}

\subsection{Localizing Partial Sums}\label{leakage_det}
\cref{ssec:layerimp} demonstrates that we can identify the high-level operations in the trace, including the layer processing and the main steps of their computation.
In \cref{ssec:gpure}, we show how we find the partial sums that leak specific weights.
In this section, we complete the profiling and identify the trace locations that leak each partial sum and, consequently, the weights.
Our approach employs TVLA~\cite{gilbert2011testing} to find statistical evidence of leakage. 
\lukasz{Commented out part about correlation analysis. }

\parhead{TVLA.}
To determine the time points in the traces where values of the partial sums $s_j$ leak, we apply fixed vs.\ random TVLA to find statistical evidence for leakage. Specifically, we collect two sets of traces.  
In the ``fixed'' set, we set the target weight to a fixed non-zero value, all other weights to zero, and all of the input to fixed randomly chosen values and collect multiple traces of performing inference with the model.
For the ``random'' set, we similarly collect multiple traces, but for each trace, we randomly select the value of the target weight.
All other weights and inputs are set to the same fixed values as in the ``fixed'' set.
We then compare the distribution of values of each time point across the two sets using Welch's $t$-test.
As common when performing side-channel attacks, if the absolute $t$-value is above 4.5, i.e., $|t|>4.5$, we mark the point as a potential leakage.\footnote{
To verify that TVLA does not yield false positives,  we check the corresponding intermediate-value correlations of a few weights to confirm the leakage. For example, when a computation followed~\cref{eq:partialsums}, we used $HD(s_{j-1}, s_j)$. All these experiments confirmed the TVLA results. 
}
\lukasz{I added a footnote. }

\begin{figure}[htb]
	\centering
	\includesvg[width=\linewidth]{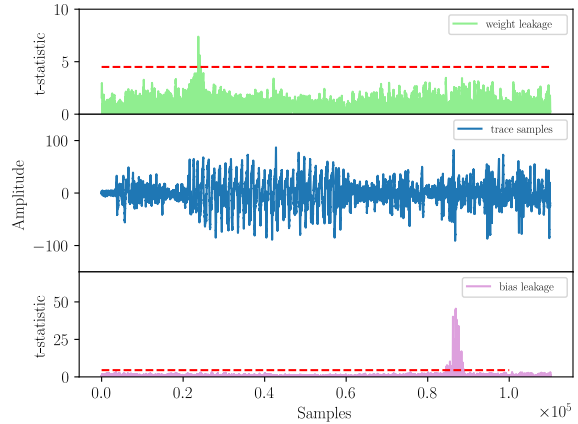}
	\caption{Result of fixed vs.\ random TVLA for the first weight (top) and the bias (bottom) in FP16 convolution with 30K and 37K traces, respectively. The middle depicts an example trace. The dashed red line indicates the 4.5 threshold. \lukasz{list amount of traces here?}\peter{fixed}}\label{fig::ex_ttest}
\end{figure}

\cref{fig::ex_ttest} shows the results of fixed vs.\ random TVLA for the first weight in the kernel in the first layer for FP16 convolution on the Jetson Nano.
Leakage is clearly evident at the start of the convolutional block, around sample 25\,000, where the $t$ value goes above 4.5.
Repeating the process for all of the weights of the kernel allows us to identify the time point for each weight at which they leak.

To create the random set for fixed vs.\ random TVLA, we need to repeatedly modify the target weight to a random value.
The TensorRT framework supports changing the weights of models, but a new CUDA context~\cite{cucontext} has to be created every time the weights of a network are changed.
This contrasts with the fixed case, where an application is not required to create a new CUDA context for every inference operation.
However, as TVLA requires changing the weights, we cannot avoid this.

\subsection{Bias Leakage}
For the bias, the final partial sum \cout or the output of the activation $c_f$ has to leak. 
Similarly, to detect leakage corresponding to the bias, we apply fixed vs.\ random bias TVLA.

\cref{fig::ex_ttest} also shows the results of TVLA for the bias of the kernel in the first layer for FP16 convolution.
A much clearer leakage is present for the bias than for the weights in the convolution operation.

After establishing where parameters leak, with the help of elastic alignment, we use static alignment on the raw traces at these points as static alignment produced higher individual TVLA peaks as well as correlation for the attack.

\subsection{CEMA Implementation in CUDA}\label{ssec:cuda}

In order to calculate correlation in CEMA, the covariances of populations have to be calculated, which can be done with two-pass algorithms.
However, for large datasets, two-pass algorithms are inefficient as the algorithm makes two iterations on the dataset.
Therefore, one-pass algorithms have been developed to estimate these statistics in large datasets~\cite{osti_1028931}.
In a one-pass algorithm, the statistics can be updated when new data points are added to the dataset.
These algorithms also make it possible to combine the statistics from subsets of the dataset to estimate the whole dataset's statistics.
This means that the statistics of each subset can be calculated in parallel, further speeding up the CEMA attack.

However, the attack has even more aspects that can be parallelized, such as the \emph{candidate} and \emph{sample} levels.
There are publicly available multithreaded implementations such as the JlSCA library~\cite{jlsca} for CPU, but these cannot fully parallelize the attack and become slow for large datasets.
Since we have a large number of candidates with FP16 weights, we decided to implement CEMA for neural networks in CUDA, parallelized on three levels: dataset, candidate, and sample.

Our implementation in CUDA launches a three-dimensional grid of thread blocks with dimensions: $(\mathit{candidates}/2, \mathit{chunks}, \mathit{samples}) = (17765, \mathit{chunks}, 32)$. 
For our use case, the number of candidates and samples are fixed at 35530 and 32, respectively.
Threads in the same warp work on the same two candidates (in parallel, due to double throughput with FP16) but correlate them with different samples.
The implementation also parallelizes CEMA
on a data set level, by setting $\mathit{chunks} > 1$.
The number of chunks can vary, but setting it to 10 already gives good results in our use case with millions of traces. 
The optimal number may be lower or higher depending on the GPU hardware.

We benchmark the multithreaded JlSCA implementation on an AMD Ryzen 7950X CPU vs.\ our CUDA implementation on a 3080 Nvidia RTX GPU.
Overall, the speedup compared to JlSCA is at least $\times5$ and typically $\times10$
when the dataset consists of millions of traces.

\section{Parameter Extraction Results}\label{sec:pex}

In this section, we demonstrate the generality of our parameter extraction framework, shown in \cref{fig::attack_proc}, on a real-world CNN architecture by targeting the baseline EfficientNet~\cite{tan2019efficientnet} to extract a kernel from its first two convolutional layers.
Note that none of the previous works have demonstrated parameter extraction from real-world CNN architectures.

In these experiments, after using the profiling information from \cref{sec:profiling}, we use the \emph{partial sums} to extract FP16 and INT8 weights on the Jetson Nano and Jetson Orin Nano, respectively. 
In addition, we also analyze the impact of batch size on the attack.

\subsection{FP16 Parameter Extraction}
Following \cite{batina2019csi},
we restrict the search space of the FP16 weights to
$[-5,5]$, as most of the parameters of trained CNNs reside in this range. To verify this, we looked at the parameters of large real-world architectures, such as our target EfficientNetB0, trained on ImageNet.
With 16-bit floats, there are 35\,330 possible candidates in this range. 

In the experiment, the parameters of the target architectures are initialized randomly.
Observe that while we attack in an iterative fashion, the trace acquisition needs to be executed only once with random known inputs. 
We target the FP16 \emph{partial sums} in each layer to extract weights and biases.

\begin{figure*}[t!]
\begin{center}
	\captionsetup[sub]{margin=16pt}
	\begin{subfigure}[t]{0.47\textwidth}
  \begin{center}
		\includesvg[height=95pt]{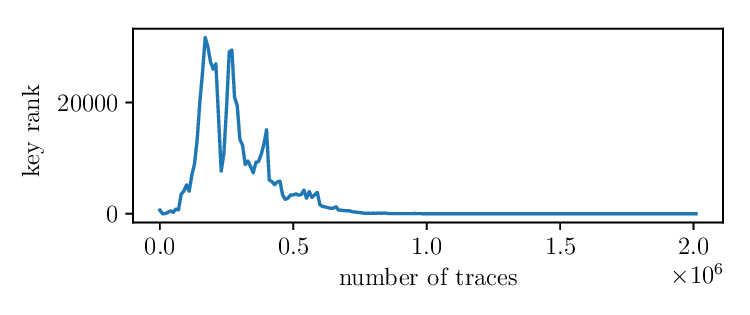}
		\caption{Key rank vs.\ number of traces using $HW$ for the third weight in the first layer with value of 0.8223.}\label{fig::kr_l1w3large}
  \end{center}
	\end{subfigure}
	\begin{subfigure}[t]{0.47\textwidth}
  \begin{center}		\includesvg[height=95pt]{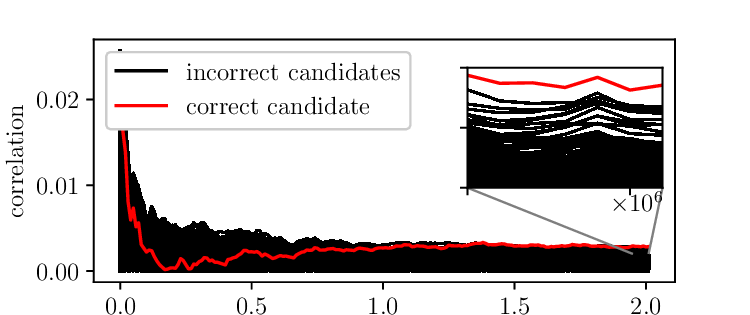}
		\caption{Correlation vs.\ number of traces using $HW$ for the third weight in the first layer with value of 0.8223.}\label{fig::correlations_l1w3large}
  \end{center}
	\end{subfigure}\\
	\begin{subfigure}[t]{0.47\textwidth}
  \begin{center}		\includesvg[height=95pt]{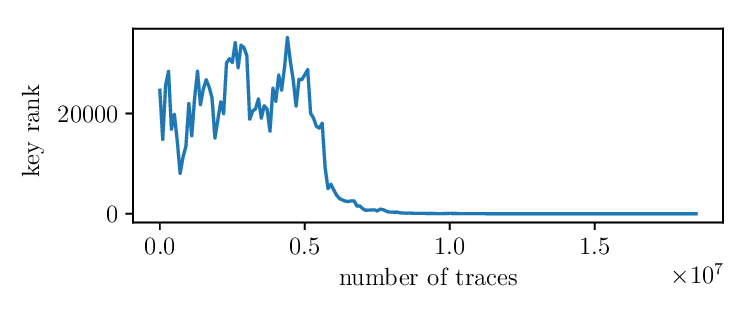}
		\caption{Key rank vs.\ number of traces using $HD$ for the ninth weight in the first layer with value of -0.7705.}\label{fig::kr_l1w9large}
  \end{center}
	\end{subfigure}
	\begin{subfigure}[t]{0.47\textwidth}
  \begin{center}		\includesvg[height=95pt]{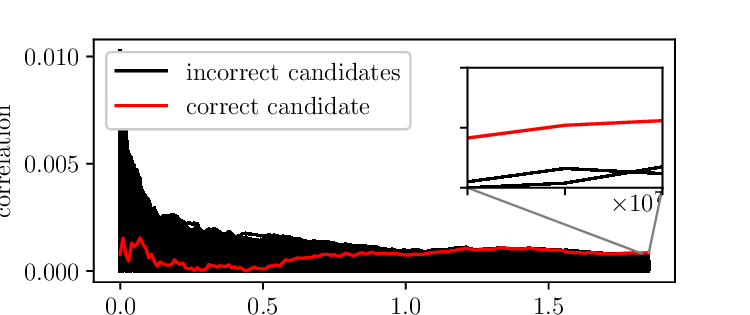}
		\caption{Correlation vs.\ number of traces using $HD$ for the ninth weight in the first layer with value of -0.7705.}\label{fig::correlations_l1w9large}
  \end{center}
	\end{subfigure}\\
	\begin{subfigure}[t]{0.47\textwidth}
  \begin{center}		\includesvg[height=95pt]{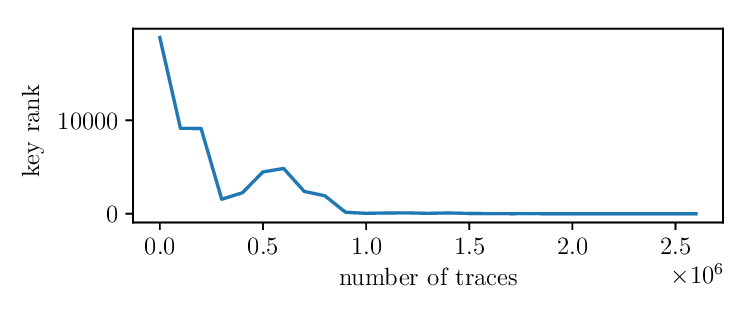}
		\caption{Key rank vs.\ number of traces for $s_{j}$ for the second weight in the second layer with the value of -0.5137.}\label{fig::kr_l2w2large}
  \end{center}
	\end{subfigure}
	\begin{subfigure}[t]{0.47\textwidth}
  \begin{center}		\includesvg[height=95pt]{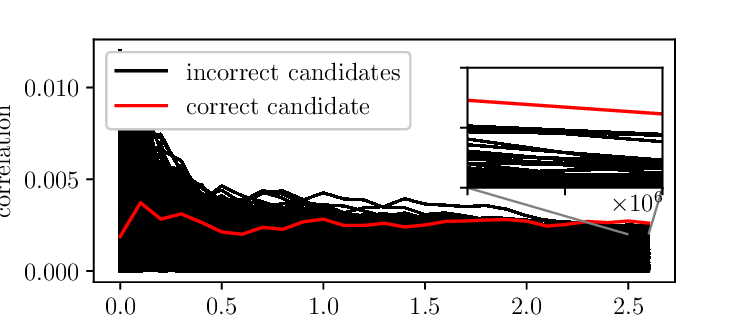}
		\caption{Correlation vs.\ number of traces for $s_{j}$ for the second weight in the second layer with the value of -0.5137.}\label{fig::correlations_l2w2large}
  \end{center}
	\end{subfigure}\\
	\begin{subfigure}[t]{0.47\textwidth}
  \begin{center}		\includesvg[height=95pt]{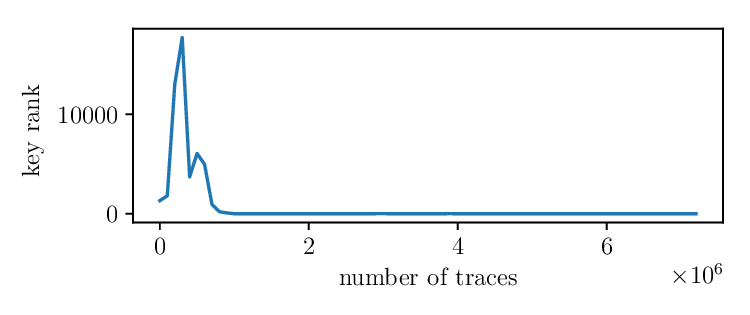}
		\caption{Key rank vs.\ number of traces for $s_{j}$ for the third weight in the second layer with the value of -0.6406.}\label{fig::kr_l2w3large}
  \end{center}
	\end{subfigure}
	\begin{subfigure}[t]{0.47\textwidth}
  \begin{center}		\includesvg[height=95pt]{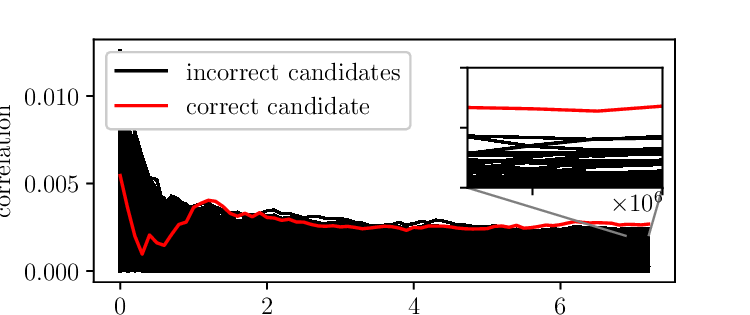}
		\caption{Correlation vs.\ number of traces for $s_{j}$ for the third weight in the second layer with the value of -0.6406.}\label{fig::correlations_l2w3large}
  \end{center}
	\end{subfigure}
\end{center}

  \caption{Key ranks and correlations of the different FP16 weights in the first and second layer on the Jetson Nano.}\label{fig::kr_corr_l12w92large}
\end{figure*}

\subsubsection{Convolutional Layer} In this experiment, we show the results of our framework
by successfully extracting FP16 parameters from the first 2 convolutional layers of the baseline EfficientNet~\cite{tan2019efficientnet} architecture by targeting the partial sums in the layer.%
\footnote{The convolutional layers in this architecture do not have biases. 
Therefore, we alter the architecture so that the first layer has biases in the kernels to demonstrate the bias extraction on this architecture as well.}
Both the HW and HD leakage models prove to be exploitable in recovering weights and biases from the layers.
Targeting the partial sums allows us to use a divide-and-conquer approach by reducing complexity, as only one FP16 weight must be extracted from a kernel at a time.

\parhead{Weight Extraction.}
For the weights, we target the partial sums $s_j$ of the 2D convolution and extract the weights of a kernel one by one.
First, we are able to extract weights using the $HW$ leakage model.
As shown in \cref{fig::kr_corr_l12w92large}\subref{fig::kr_l1w3large} and \cref{fig::kr_corr_l12w92large}\subref{fig::correlations_l1w3large},
we successfully recover the third weight of a kernel in the first layer.
For the second layer, \cref{fig::kr_corr_l12w92large}\subref{fig::kr_l2w2large} and \cref{fig::kr_corr_l12w92large}\subref{fig::correlations_l2w2large} show the key ranking and correlation for the second weight in the second layer.
In addition, \cref{fig::kr_corr_l12w92large}\subref{fig::kr_l2w3large} and \cref{fig::kr_corr_l12w92large}\subref{fig::correlations_l2w3large} show the key ranking and correlation for the third weight in the second layer, demonstrating that our attack extends to larger layers as well.

The $HD(s_{j-1}, s_j)$ leakage model targeting a register update can also be exploited to recover weights: \cref{fig::kr_corr_l12w92large}\subref{fig::kr_l1w9large} and \cref{fig::kr_corr_l12w92large}\subref{fig::correlations_l1w9large} show the key rankings and correlations for the 9th weight in the first layer of the large architecture using HD. The attack works similarly for the other weights.
As shown in \cref{fig::kr_corr_l12w92large}\subref{fig::kr_l1w9large}, the convergence behavior for HD
is different from that for HW as it starts to converge slower.

\begin{figure*}[t!]
\begin{center}
	\captionsetup[sub]{margin=16pt}
	\begin{subfigure}[t]{0.47\textwidth}
   \begin{center}
		\includesvg[height=95pt]{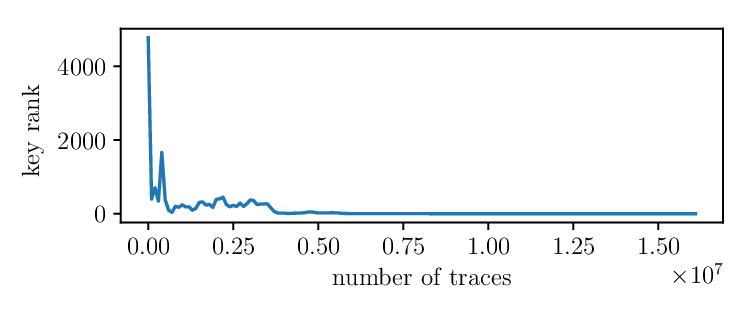}
		\caption{Key rank vs.\ number of traces for ReLU.
		}\label{fig::key_rank_ReLU}
   \end{center}
	\end{subfigure}
	\begin{subfigure}[t]{0.47\textwidth}
   \begin{center}
		\includesvg[height=95pt]{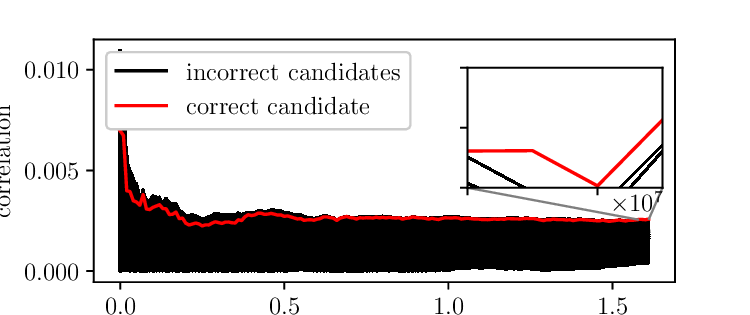}
		\caption{Correlation vs.\ number of traces for ReLU.
		}\label{fig::correlations_ReLU}
   \end{center}
	\end{subfigure}\\
	\begin{subfigure}[t]{0.47\textwidth}
   \begin{center}
		\includesvg[height=95pt]{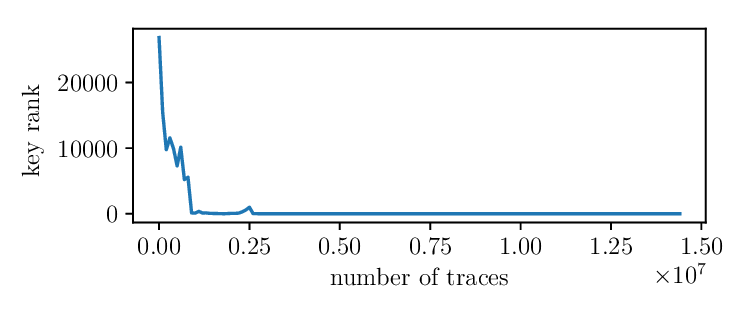}
		\caption{Key rank vs.\ number of traces for $c_{out}$.}\label{fig::key_rank_bias2}
   \end{center}
	\end{subfigure}
	\begin{subfigure}[t]{0.47\textwidth}
   \begin{center}
		\includesvg[height=95pt]{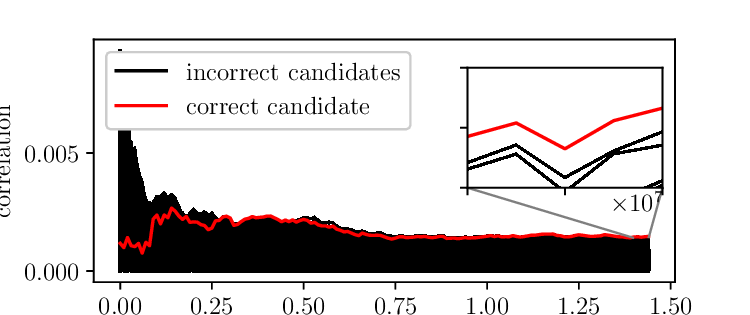}
		\caption{Correlation vs.\ number of traces
			for $c_{out}$.}\label{fig::correlations_bias2}
   \end{center}
	\end{subfigure}\\
	\begin{subfigure}[t]{0.47\textwidth}
   \begin{center}
		\includesvg[height=95pt]{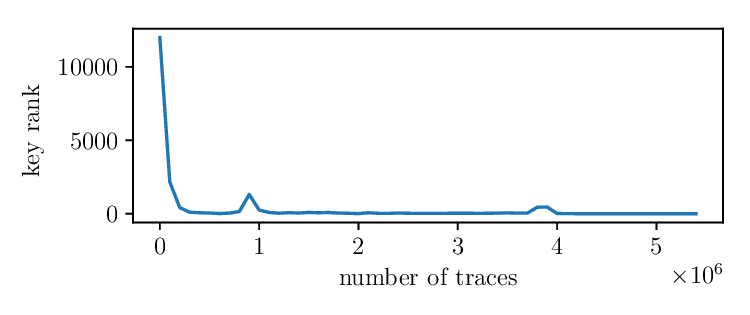}
		\caption{Key rank vs.\ number of traces when $c_{sum}$ is overwritten with $c_{out}$.}\label{fig::key_rank_bias1HDlarge}
   \end{center}
	\end{subfigure}
	\begin{subfigure}[t]{0.47\textwidth}
   \begin{center}
		\includesvg[height=95pt]{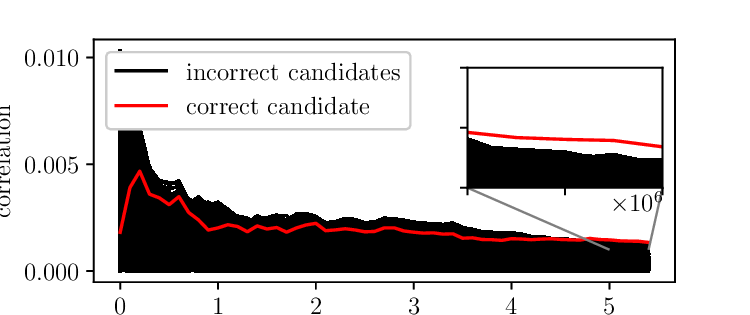}
		\caption{Correlation vs.\ number of traces when $c_{sum}$ is overwritten with $c_{out}$.}\label{fig::correlations_bias1HDlarge}
   \end{center}
	\end{subfigure}

	\caption{Key ranks and correlations of the FP16 biases with different leakage models on the Jetson Nano.}\label{fig::kr_corr_bias_l12HWHDsmalllarge}
\end{center}
\end{figure*}

\begin{figure*}[t!]
\begin{center}

	\captionsetup[sub]{margin=16pt}
	\begin{subfigure}[b]{0.47\textwidth}
   \begin{center}
		\includesvg[height=95pt]{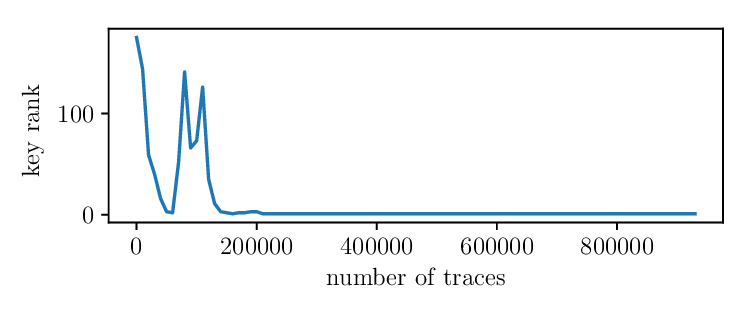}
		\caption{Key rank vs.\ number of traces with HD leakage model for the fourth weight in the convolutional kernel.}\label{fig::key_rank_w4_conv_orin_batch1}
   \end{center}
	\end{subfigure}
	\begin{subfigure}[b]{0.47\textwidth}
   \begin{center}
		\includesvg[height=95pt]{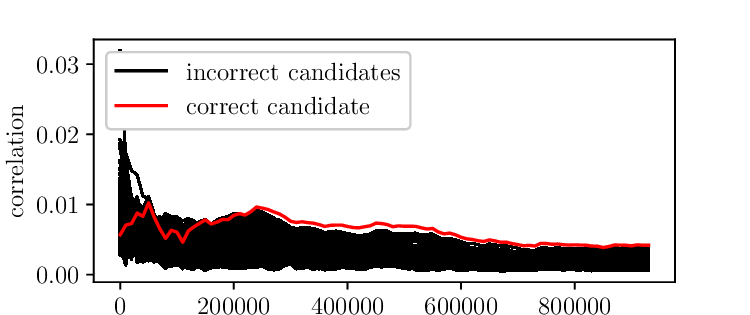}
		\caption{Correlation vs.\ number of traces with HD leakage model for the fourth weight in the convolutional kernel.}\label{fig::correlations_w4_conv_orin_batch1}
   \end{center}
	\end{subfigure}\\
	\begin{subfigure}[b]{0.47\textwidth}
   \begin{center}
		\includesvg[height=95pt]{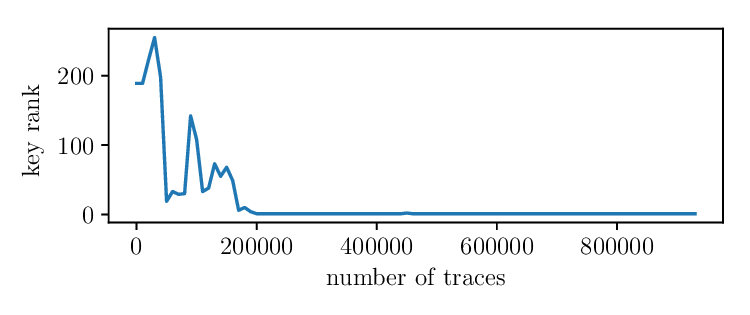}
		\caption{Key rank vs.\ number of traces with HD leakage model for the fifth weight in the convolutional kernel.}\label{fig::key_rank_w5_conv_orin_batch1}
   \end{center}
	\end{subfigure}
	\begin{subfigure}[b]{0.47\textwidth}
   \begin{center}
		\includesvg[height=95pt]{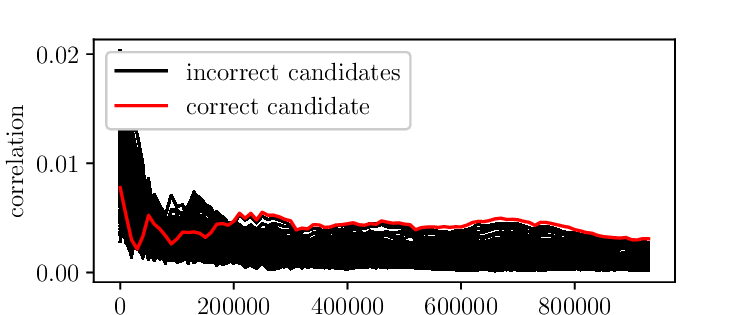}
		\caption{Correlation vs.\ number of traces with HD leakage model for the fifth weight in the convolutional kernel.}\label{fig::correlations_w5_conv_orin_batch1}
   \end{center}
	\end{subfigure}\\
	\begin{subfigure}[b]{0.47\textwidth}
   \begin{center}
		\includesvg[height=95pt]{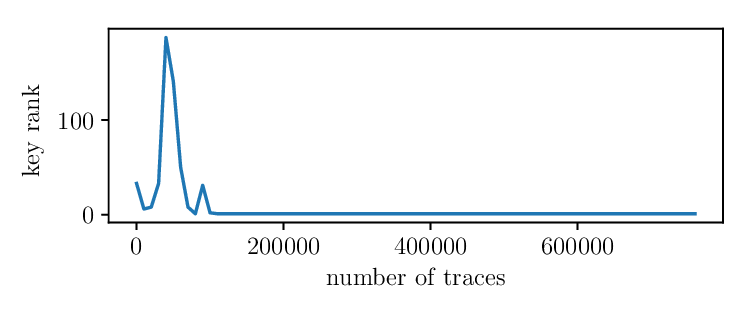}
		\caption{Key rank vs.\ number of traces with HW leakage model for the 8th weight in the dense layer.}\label{fig::key_rank_HW_w8_dense_orin_batch1}
   \end{center}
	\end{subfigure}
	\begin{subfigure}[b]{0.47\textwidth}
   \begin{center}
		\includesvg[height=95pt]{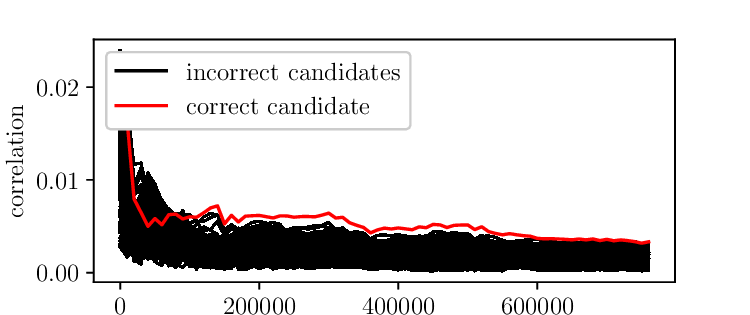}
		\caption{Correlation vs.\ number of traces with HW leakage model for the 8th weight in the dense layer.}\label{fig::correlations_HW_w8_dense_orin_batch1}
   \end{center}
	\end{subfigure}\\
	\begin{subfigure}[b]{0.47\textwidth}
   \begin{center}
		\includesvg[height=95pt]{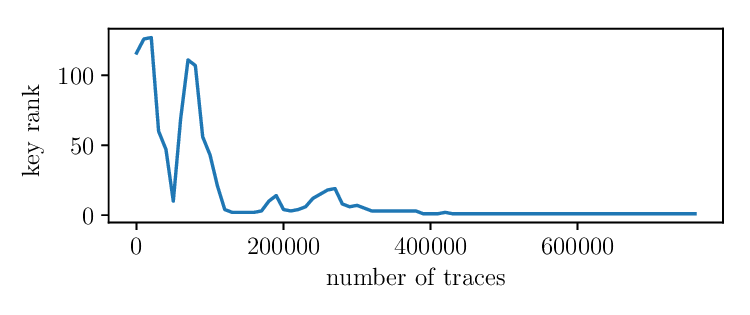}
		\caption{Key rank vs.\ number of traces with HD leakage model for the 8th weight in the dense layer.}\label{fig::key_rank_HD_w8_dense_orin_batch1}
   \end{center}
	\end{subfigure}
	\begin{subfigure}[b]{0.47\textwidth}
   \begin{center}
		\includesvg[height=95pt]{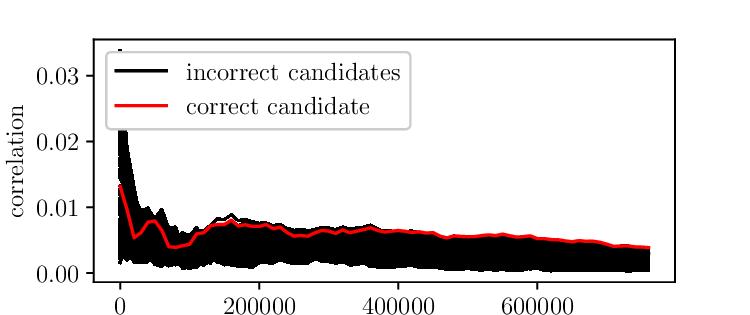}
		\caption{Correlation vs.\ number of traces with HD leakage model for the 8th weight in the dense layer.}\label{fig::correlations_HD_w8_dense_orin_batch1}
   \end{center}
	\end{subfigure}
\end{center}

	\caption{Key ranks and correlations of the different INT8 weights of convolutional and dense layers on the Jetson Orin Nano.}\label{fig::kr_corr_l1_orin_conv_batch1}
\end{figure*}

\parhead{Bias Extraction.} Similarly to the weights, we are also able to use $HD$ leakage model to extract the bias: we targeted the register update from $c_{\mathit{sum}}$ to $c_{\mathit{out}}$: $\mathit{HD}(c_{\mathit{sum}}, c_{\mathit{out}})$.
\cref{fig::key_rank_bias1HDlarge} and \cref{fig::correlations_bias1HDlarge} show the results for the bias in the first layer.
The key rank drops quickly and converges to key rank 0 in 5 million traces. 
Overall, there is a significant variance in the number of required traces to recover individual weights and biases, but an upper bound of 20 million traces proves sufficient in our experiments.

\subsection{INT8 Parameter Extraction}

For INT8 weights, we demonstrate our parameter extraction framework on the first convolutional layer with the same EfficientNetB0 configuration as in the FP16 case,
also with different batch sizes. Similarly, we target the partial sums in the INT8 convolution to extract the weights.
In addition, we provide results on dense layer parameter extraction with INT8 weights.

\subsubsection{Convolutional Layer}

\parhead{Weight Extraction.}
In this case, the partial sums can depend on multiple weights, depending on the number of input channels.
In this experiment, the number of input channels to the first layer is one, so the partial sums depend on exactly one 8-bit weight.
If the number of input channels is larger than 1, as is the case in subsequent layers, then the target intermediate values depend on multiple weights, increasing the complexity of the attack.

Similarly to the FP16 case, we extract the weights in a kernel one by one by targeting the partial sums in the convolution.
\cref{fig::kr_corr_l1_orin_conv_batch1}\subref{fig::key_rank_w4_conv_orin_batch1} and \cref{fig::kr_corr_l1_orin_conv_batch1}\subref{fig::key_rank_w4_conv_orin_batch1}
show the results for 4th weight in a kernel, while \cref{fig::kr_corr_l1_orin_conv_batch1}\subref{fig::key_rank_w5_conv_orin_batch1} and \cref{fig::kr_corr_l1_orin_conv_batch1}\subref{fig::correlations_w5_conv_orin_batch1}
show the results for the 5th weight in a kernel, respectively. In our experiments, on average, 300K traces were enough for the correct key candidates to reach key rank 0.
Both the HW and HD leakage models are exploitable to recover individual weights.    

\parhead{Impact of Batch Size.}
The batch size does influence the number of traces required to extract weights but not significantly, as 500K traces are sufficient to reach key rank 0. 
The difference between smaller and larger batch sizes is the number of executing threads.
For instance, the first convolutional layer is executed with 512 and 7\,232 threads for batch sizes of 1 and 16, respectively.
Even though there are 14 times more threads for batch size 16, the number of executing threads physically in parallel is limited, e.g., the warp schedulers still only issue instructions for at most 32 threads at a time.
Therefore, on smaller GPUs, the main impact of batch size might be only the linear increase in execution time.
On the other hand, a larger batch size provides even more partial sums in one trace, and a horizontal attack~\cite{DBLP:conf/icics/ClavierFGRV10} exploiting this might enhance the efficacy of the attack. We leave analyzing this for future work.

\subsubsection{Dense Layer}
We present parameter extraction results of a neural network containing a single dense layer with INT8 weights running on the Jetson Orin Nano. In the experiments, our target dense layer has 512 nodes, and the input size $s$ of the layer is 784. 
Therefore, each node has 784 weights associated with it. 
Our leakage modeling experiments for dense layers discovered that, similarly to convolutional layers, the HW and HD leakage models are both viable for mounting a successful parameter extraction attack.
However, these exploitable leakage models depend on multiple, not just one, 8-bit weights because the input size to dense layers is significantly larger than 4 in real-world applications. Therefore, most or all partial sum results depend on four weights, i.e., we face a complexity of 32 bits with the INT8 data type. 
To combat this issue, we apply the chosen-input attack mentioned earlier by setting three channels to 0 so that the registers that hold the input values contain only 1 non-zero input. 
Therefore, the final result will only depend on one 8-bit weight. 
Since all the nodes in the layer receive the same inputs, the chosen input attack does not require a new trace-set for every node separately. 
Therefore, the attack is \emph{independent} of the size of the layer.

\cref{fig::kr_corr_l1_orin_conv_batch1}\subref{fig::key_rank_HW_w8_dense_orin_batch1} and \cref{fig::kr_corr_l1_orin_conv_batch1}\subref{fig::correlations_HW_w8_dense_orin_batch1}
show the results with HW leakage model while \cref{fig::kr_corr_l1_orin_conv_batch1}\subref{fig::key_rank_HD_w8_dense_orin_batch1} and \cref{fig::kr_corr_l1_orin_conv_batch1}\subref{fig::correlations_HD_w8_dense_orin_batch1}
show the results with HD leakage model for the 8th weight in the first node. 
In our experiments, on average, 300K traces were enough for the correct key candidates to reach key rank 0, similar to the convolutional layer.
In addition, both the HW and HD leakage models are exploitable to recover individual weights, which is unsurprising since the same instruction is used for the layers.

\subsubsection{Attacking Deeper Layers.}
In deeper layers with INT8 weights, each partial sum depends on four 8-bit weights as the number of input channels is larger than four. Generally, this increases the complexity of the attack to guessing four weights at a time, so $2^{32}$ candidates, which is computationally expensive. However, if the attacker has enough computing resources, the attack can still be mounted by attacking 32 bits. 

However, the complexity can be reduced in two similar ways:
\begin{enumerate}[nosep, leftmargin=*]
	\item Collect traces with random inputs and choose only those where the inputs to deeper layers in specific channels are zero.
	\item Collect traces with chosen inputs so deeper layers receive zero inputs in specific channels.
\end{enumerate}

\parhead{Random Inputs.} This approach implies that the attacker has to collect significantly more traces to have 300K traces with the desired 0-value inputs in specific channels.
However, the inputs in DNNs are usually centered around zero, and some of their quantized values are exactly 0. More importantly, the ReLU activation function greatly enhances sparsity in the inputs due to setting negative values to 0.

\parhead{Chosen Inputs.} In this approach, the attacker has to solve systems of linear equations to generate inputs to the first layer such that the inputs to deeper layers are 0, similarly to~\cite{Gongye2023:SCA-DPU}. 
The number of collected traces is significantly less than that of random inputs, but whether this approach works for deeper layers in practice is still to be determined.
Nevertheless, quantization and the ReLU activation function also enhance this approach as these factors make the solutions space larger. 

\subsubsection{Parameter extraction comparisons}

\begin{table}[t]
	\centering
	\begin{tabular}{lll}
		\toprule
		                 & Jetson Nano & Jetson Orin Nano \\ [0.5ex]
		\midrule
		weights data type & FP16 & INT8 \\
		bias data type & FP16 & FP32 \\
		\# of extracted layers     & 2         & 1             \\
		\# of extracted parameters   & 9564      & 316 \\          
		bias extraction & \checkmark & -  \\
		\# of max req. traces & 20M & 3M \\
		\bottomrule
	\end{tabular}
	\caption{Parameter extraction results for different GPUs for the EfficientNetB0 architecture.}\label{table:param_ex_result}
\end{table}

The summary of parameter extraction of the target EfficientNetB0 architecture on different GPUs is presented in \cref{table:param_ex_result}.
For the Jetson Nano with FP16 parameters, the first two convolutional layers can be extracted with a complexity of $2^{16}$ for each parameter. However, a full model extraction is still computationally expensive as the model contains millions of parameters.
In this case, the parameters required at most 20M traces for successful extraction. 
Observe that the number of traces should not increase even if we attack more layers because the attack does not make assumptions about layers' inputs. 

For the Jetson Orin Nano with INT8 weights, we were only able to extract the weights of the first layer as the parameters in the second layer have a complexity of $2^{32}$ candidates due to how partial sums depend on 4 weights. In addition, the biases are represented as 32-bit floats that also provide a complexity of $2^{32}$ candidates.
Although the number of required traces is at most 3M for the first layer weights, subsequent layers require most likely more as there are $2^{32}$ candidates.
Overall, even with FP16 parameters, a full model extraction requires computing resources that are beyond our capabilities but not beyond a well-founded commercial or state attacker.

\section{Discussion}\label{sec:disc}

\subsection{Approaching Additional platforms}

Our approach is demonstrated on CUDA-enabled GPUs, but we expect that our methodology 
is applicable to other platforms as well.
An attack on another 
platform would also start with a profiling phase where the target architecture's implementation is reverse-engineered on a low level.
This is necessary to identify and localize partial sums allowing for parameter extraction.
After the profiling phase, the attacker is equipped with the appropriate leakage models and locations to extract the parameters. 
We expect that there would be some differences with respect to the exact intermediate values that would need to be targeted, but the approach should be similar.

\subsection{Desktop/Datacenter GPUs}
Against large GPUs,
our attack can be extended and is likely to be more expensive depending on the target neural network’s size.
One key difference between the Jetsons and desktop/datacenter GPUs is the number of streaming multiprocessors. 
Ideally, for a large GPU, an attacker would first locate all the SMs of the target GPU. Afterward, each SM could be scanned with an EM probe to see if there is any activity of interest.
If the target neural network is extremely large and saturates all the SMs during inference, multiple probes may be used to cover all of them and collect traces in parallel for each SM.
Therefore, the equipment and overall cost is higher the more SMs the target GPU has.

\subsection{Limitations}
While we demonstrated parameter extraction on multiple GPUs, the coverage is still limited: 

\parhead{Concurrent Applications.} GPUs are able to handle and schedule concurrently CUDA functions from multiple applications.
This
might introduce noise, but
it depends on the required resources for each CUDA function.
Suppose a CUDA function takes up most of the GPU resources (e.g., shared memory, registers, etc.). In that case, a different CUDA function will only be scheduled after all the thread blocks
in the previous CUDA function finished execution~\cite{cuda_concurrency}. Therefore, on GPUs with few SMs, this is less likely to be an issue as the GPU's resources are already saturated due to the large computational demand of DNN inference.

\parhead{Bias Extraction in INT8 Implementations}
We demonstrated the extraction of biases from convolutional layers for FP16 implementations but not INT8 implementations. 
These implementations only use INT8 for the weights but not for the bias. The bias in these implementations has FP32 data type, which is beyond our computing capabilities to extract.

\subsection{Mitigation}
Traditional ways to contain electromagnetic emanation, such as proper shielding or introducing noise to decrease the Signal-to-Noise ratio, could alleviate the problem~\cite{mangard2008power}.
Specifically against parameter extraction, one of the possible countermeasures, which is also mentioned in the CSI-NN paper~\cite{batina2019csi}, is shuffling~\cite{veyrat2012shuffling} the order of multiplications in the layers, which can make it significantly harder for an adversary to recover the weights.
Additionally, masking~\cite{coron2000boolean, prouff2013masking}
can also decouple the side-channel measurements and the processed data.
However, this comes at the price of execution speed, which might not be desired in real-time systems.
Specifically for convolution, the registers containing the results of the partial sums can be initialized with the bias of the kernel instead of initializing them with zeros.
This would prompt an adversary to mount a CEMA attack where the correct $b+w_{1}$ pair has to be recovered first. The complexity of this attack would be 32 bits due to 16 bits of complexity for the weight and bias separately in the FP16 case.
However, in the INT8 case, the bias is a single-precision float, so it cannot be used to initialize the accumulator registers. 

\begin{table}[htb]
 \footnotesize
	\centering
	\begin{tabular}{@{}lllll@{}}
		\toprule
		&         & Clock             & Side & Parameter  \\ [0.5ex]
		 Author&  Platform   &  (MHz)              & channel &  datatype \\ [0.5ex]
		\midrule
		Batina, et al.~\cite{batina2019csi}                          & MCU & 20, 84                         & EM           & FP32           \\
		Dubet, et al.~\cite{Dubey2020:Maskednet}                     & FPGA            & 24                             & Power        & Binary                \\
		Yoshida, et al.~\cite{Yoshida2020:ModelReverseEngineering}   & FPGA            & 25                             & Power        & INT8                \\
		Regazzoni, et al.~\cite{Regazzoni2020:MLHardwareSecurity}    & FPGA            & N/A\footnotemark               & EM           & Binary                \\
		Yli-M{\"a}yry, et al.~\cite{Yli2021:ExtractionBNN} & FPGA            & N/A\footnotemark[\thefootnote] & EM           & Binary                \\
		Li, et al.~\cite{Li2022:PowerAttacksDNN}                     & FPGA            & 25                             & Power        & INT8                \\
		Joud, et al.~\cite{Joud2022:PracticalSCADNN}                 & MCU & 100                            & EM           & FP32               \\
		Gongye et al.~\cite{Gongye2023:SCA-DPU}                      & FPGA            & 320                            & EM           & INT8                \\
		\textbf{BarraCUDA}                                           & \textbf{GPU}    & \textbf{625}, \textbf{920}     & EM           & INT8, FP16            \\
		\bottomrule
	\end{tabular}
	\caption{Comparison with related work.}\label{table:rw}
\end{table}

\subsection{Related Work}

To the best of our knowledge, no previous work has been able to extract the parameters of neural networks on GPU using physical side-channel.
Previous works have demonstrated parameter extraction on microcontrollers and FPGAs using power or EM side channel, as shown in \cref{table:rw}.
In addition, these attacks were performed on neural networks with binary parameters~\cite{Dubey2020:Maskednet,Regazzoni2020:MLHardwareSecurity,Yli2021:ExtractionBNN}, 8-bit parameters~\cite{batina2019csi, Yoshida2020:ModelReverseEngineering,Li2022:PowerAttacksDNN,Gongye2023:SCA-DPU} or 32-bit parameters~\cite{batina2019csi,Joud2022:PracticalSCADNN}.
Our work demonstrates parameter extraction of 8- and 16-bit parameters.
Furthermore, our work presents a CEMA attack on weights where the number of cores and the clock frequency at these cores operate are significantly larger than in related works.
The large number of cores, with almost 1GHz clock frequency, presents a challenge in both the measurement and attack stages.
Given that GPUs are the backbone of AI, it is of utmost importance to assess the resilience of GPU accelerated workloads against weight extraction attacks, a task our research addresses.

\footnotetext{The clock frequency is not disclosed in these attacks, but it is at most 800MHz as both attack XILINX ZYNQ chip~\cite{zynq_datasheet}.\yval{Make sure this appears on the same page as Table 1}}

%

\section{Conclusions}\label{sec:conc}

In this work, we analyzed the GPUs of Nvidia Jetson Nano and Nvidia Jetson Orin Nano, commonly chosen platforms for real-world neural network implementations, for resilience against side-channel attacks that aim to extract the weights of the target NN.
First, we find multiple vulnerable points where the GPUs leak information about the parameters of the target DNN.
Subsequently, we demonstrate the extraction of weights and biases of convolutional and dense layers.
Overall, the neural network implementations of Nvidia's TensorRT framework are vulnerable to parameter extraction using
EM
side-channel attack despite the networks running in a highly parallel and noisy environment. 
Protecting their implementations in security or privacy-sensitive applications remains an open problem.

\ifAnon\else
\section*{Acknowledgments}

This research was supported by:
an ARC Discovery Project number DP210102670; 
the Deutsche Forschungsgemeinschaft (DFG, German Research Foundation) under Germany's Excellence Strategy - EXC 2092 CASA - 390781972; 
Ai-SecTools (VJ02010010); 
PROACT project of Dutch Research Agenda (NWA.1215.18.014) and Netherlands Organisation for Scientific Research (NWO); TTW PREDATOR project 19782 (NWO). 

\fi
\bibliographystyle{plain}
\bibliography{main.bib}

\appendix
\section{Convolution implementation details}\label{sec:convimpdetails}

\begin{lstlisting}[caption=FP16 convolution code snippet, label=lst:fp16, float=htb]
HFMA2     R0, R94, R109, R0;
HFMA2     R3, R92, R108, R3;    
LDS.U.128 R116, [R88+0x210];
\end{lstlisting}

\begin{figure}[htb]
	\centering
	\includesvg[width=.4\textwidth]{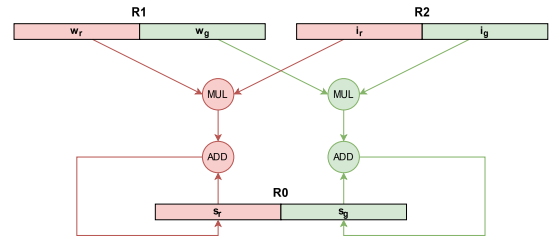}
	\caption{HFMA2 instruction operation with two input channels.}\label{fig::hfma2}
\end{figure}

In this section, we show how the computation of partial sums differs based on the representation of the parameters and how this influences the attack.
Specifically, the FP16 and INT8 data types have their own specialized instructions where the GPU registers are akin to vector registers.

\parhead{FP16 convolution.}
As demonstrated in \cref{lst:fp16}, FP16-based convolution uses the \instr{HFMA2} instruction, which performs two half-precision fused-multiply-adds in parallel.
As illustrated in \cref{fig::hfma2}, the instruction takes three input registers, each is a two-lane vector.
It multiplies the values in the matching channels of two registers, adds the result to the matching lanes of the third, and stores the result in the output register.
In all instances we have seen, a single register is used as an accumulator, where the multiplication result is added to it. 
Our leakage model targets each partial sum (16 bits) that is written into this accumulator register.

\begin{lstlisting}[caption=INT8 convolution code snippet, label=lst:int8, float=htb]
IDP.4A.S8.S8 R62, R71, R83, R62; 
IDP.4A.S8.S8 R59, R69, R82, R59;    
LDS.128      R64, [R97+0x200];
\end{lstlisting}

\begin{figure}[htb]
	\centering
	\includesvg[width=.4\textwidth]{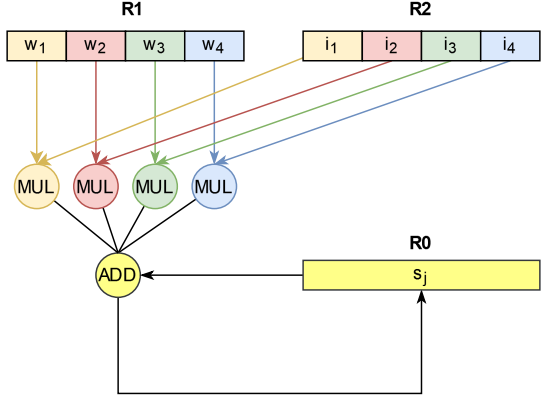}
	\caption{IDP.4A instruction operation with 4 input channels.}\label{fig::idp4a}
\end{figure}

\parhead{INT8 convolution.}
\cref{lst:int8} shows an example of code used in the INT8 implementation.
The implementation uses the \instr{IDP.4A} instruction to perform a 4-way dot product and accumulate operation, depicted in \cref{fig::idp4a}.
The instruction first multiplies the elements in matching lanes in two registers.
It then sums the results and adds them to a third register, which stores a signed 32-bit integer.
As in the FP16 case, in all uses we have seen, \instr{IDP.4A} uses a single register as input and output numbers.

To extract weights, our leakage model targets the 32-bit partial sum ($s_c$ in \cref{fig::idp4a}) stored in the accumulator register. 
As the result depends on all four input lanes, the complexity of the CEMA attack grows to up to 32 bits.
In practice, the code often uses less than four lanes. 
Specifically, when processing greyscale images, convolutions tend to use only one lane, whereas when processing color images, convolutions use three lanes, one for each color channel.

Using the INT8 representation on the Orin Nano, Tensor Core implementations use the \instr{IMMA} (Integer Matrix Multiply and Accumulate) instruction. 
\instr{IMMA} works on a warp level, meaning the threads in the warp must cooperate, but its execution is similar to \instr{IDP.4A} on a thread level. 
The implementations we experiment with in this paper do not use the \instr{IMMA} instruction. 
However, the techniques we develop for parameter extraction are directly applicable to it as well.

\parhead{Dense layer.}
The implementation of dense layers follows the same design as convolutional layers.
In these implementations, each accumulator register holds the partial sums of the weighted sum of a node in the layer.
Similarly to INT8 convolutional implementations, the dense layer implementation uses the \instr{IDP.4A} instruction to calculate the partial sums. 
Consequently, as in the convolutional, there is a need to guess 32 bits, except in edge cases, where some channels are set to zero.

\end{document}